\documentclass{article}

\input{tcilatex}
\begin{document}

\title{Symmetric form of governing equations\\
for capillary fluids}
\author{Sergey Gavrilyuk\thanks{
sergey.gavrilyuk@univ-cezanne.fr} \ and Henri Gouin\thanks{
henri.gouin@univ-cezanne.fr }}
\date{{\small Laboratoire de Mod{\'e}lisation en M{\'e}canique et
Thermodynamique, EA2596, Universit{\'e} d'Aix-Marseille, 13397 Marseille
Cedex 20, France}}
\maketitle

\begin{abstract}
In classical continuum mechanics, quasi-linear systems of conservation laws
can be symmetrized if they admit an additional convex conservation law. In
particular, this implies the hyperbolicity of governing equations. For
capillary fluids, the internal energy depends not only on the density but
also on its derivatives with respect to space variables. Consequently, the
governing equations belong to the class of dispersive systems. In that case
we propose a symmetric form of governing equations which is different from
the classical Godunov - Friedrichs - Lax representation. This new symmetric
form implies the stability of constant solutions.
\end{abstract}

\section{Introduction}

Quasi-linear systems of conservation laws can be symmetrized, if they admit
an additional \emph{convex} conservation law (Godunov, 1961, Friedrichs and
Lax, 1971). The symmetric form implies hyperbolicity of governing equations.
For conservation laws with vanishing right-hand side, the hyperbolicity is
equivalent to stability of constant solutions with respect to perturbations
of the form $\ e^{i(\mathbf{k}^{\ast }\mathbf{x}-\lambda t)},\ i^{2}=-1,\
\mathbf{k}^{\ast }=(k_{1},\cdots ,k_{n}),\ \mathbf{x}=(x^{1},\cdots ,x^{n})$%
, where $^{"\ast"}$ denotes the transposition. Indeed, the following
symmetric form of governing equations for an unknown vector variable $%
\mathbf{v}$
\begin{equation}
A\frac{\partial \mathbf{v}}{\partial t}+\sum_{i=1}^{n}B^{i}\frac{\partial
\mathbf{v}}{\partial x^{i}}=0
\end{equation}%
where matrix $A=A^{\ast }$ is positive definite, $B^{i}=(B^{i})^{\ast }$,
implies the dispersion relation
\begin{equation}
\mathtt{det}(B-\lambda A)=0,\quad B=\sum_{i=1}^{n}B^{i}k_{i}
\end{equation}%
which determines real values of $\lambda $ for any real wave vector $\mathbf{%
k}$.

In this note we get an analog of symmetric form (1) for \emph{\ }equations
of \emph{capillary fluids} that belong to the class of \emph{dispersive
systems}, because the internal energy depends not only on the density but
also on its derivatives with respect to space variables. We will see that
the analog of equation (2) is
\begin{equation}
\mathtt{det}(B+iC-\lambda A)=0
\end{equation}%
where $C=-C^{\ast }$ is an antisymmetric matrix depending on the wave vector
$\mathbf{k}$. Since $B+iC$ is Hermitian matrix and the symmetric matrix $A$
is positive definite, all the frequencies $\lambda $ are also real. For a
capillary fluid the matrix $C$ is of the form
\[
C=-C^{\ast }=\left(
\begin{array}{ccc}
0 & \mathbf{0}^{\ast } & \mathbf{0}^{\ast } \\
\mathbf{0} & \mathit{O} & -\rho _{e}\mathbf{kk}^{\ast } \\
\mathbf{0} & \rho _{e}\mathbf{kk}^{\ast } & \mathit{O}%
\end{array}%
\right)
\]%
where $\rho _{e}$ is the equilibrium fluid density and $\mathit{O}$ is the
zero-matrix $3\times 3$.

Here and later, for any vectors $\mathbf{a,b}$ we use the notation $\mathbf{a%
}^{\ast }\mathbf{b}$ for the scalar product (the line is multiplied by the
column vector) and $\mathbf{a}^{\ }\mathbf{b}^{\ast }$ for the tensor
product (or $\mathbf{a}\otimes \mathbf{b}$ the column vector is multiplied
by the line vector). Divergence of a linear transformation $A$ is the
covector $\func{div}(A)$ such that, for any constant vector $\mathbf{a}$, $%
\func{div}(A)$ $\mathbf{a}=$ $\func{div}(A\mathbf{a})$. The identical
transformation is denoted by $I$.

In section 2 we present the multi-dimensional case in Eulerian coordinates
for a particular form of the internal energy. In section 3 we consider in
Lagrangian coordinates a one-dimensional case for the general form of
internal energy.

\section{Governing equations in Eulerian coordinates}

The internal energy per unit volume of a capillary fluid is taken in the
form
\begin{equation}
e(\rho ,\eta ,\mathbf{w)=\varepsilon (}\rho ,\eta \mathbf{)+}\frac{%
c\left\vert \mathbf{w}\right\vert ^{2}}{2}
\end{equation}%
where $\rho $ is the fluid density, $\mathbf{w}= \rm{grad}\, \rho$ (or $\mathbf{w}^{\ast }=\nabla \rho $), $\ \eta $
is the entropy per unit volume, $c$ is the \emph{capillarity coefficient}
which is assumed to be a constant [Rocard, 1952, Rowlinson and Widom, 1984].
The \emph{homogeneous} energy $\varepsilon $\ satisfies the Gibbs identity%
\begin{equation}
d\varepsilon =\mu \ d\rho +\theta \ d\eta
\end{equation}%
where $\mu =(\varepsilon +P-\theta \eta )/\rho $ is the chemical potential, $%
\theta $ is the temperature, $P=\displaystyle\rho \frac{\partial \varepsilon
}{\partial \rho }-\varepsilon $ is the thermodynamic pressure. By using
Hamilton's principle, governing equations of such a fluid were obtained by
Casal (1972) (see also Casal and Gouin, 1985, Gavrilyuk and Shugrin, 1996).
They are in the form%
\begin{equation}
\begin{array}{c}
\rho _{t}+\func{div}\ \mathbf{j}=0 \\
\eta _{t}+\func{div}\left(\displaystyle\frac{\eta }{\rho }\ \mathbf{j}%
\right)=0 \\
\mathbf{j}_{\,t}^{\ast }+\func{div}\left( \displaystyle\frac{\mathbf{jj}%
^{\ast }}{\rho }+p\,I+c\mathbf{ww}^{\ast }\right) =\mathbf{0}^{\ast }%
\end{array}%
\end{equation}%
where $\mathbf{j}=\rho \mathbf{u}$, $\mathbf{u}$ is the velocity vector,
index $t$ is the partial derivative with respect to time, $\displaystyle %
p=\rho \frac{\delta \varepsilon }{\delta \rho }-e=P-c\left( \rho \,\func{div}%
\mathbf{w}+\frac{\left\vert \mathbf{w}\right\vert ^{2}}{2}\right) $, where $%
\displaystyle\frac{\delta }{\delta \rho }$ means the variational derivative
with respect to $\rho $. By using (4) and (5) we can obtain from (6) the
energy conservation law
\begin{equation}
\left( \varepsilon +\frac{\left\vert \mathbf{j}\right\vert ^{2}}{2\rho }+%
\frac{c\left\vert \mathbf{w}\right\vert ^{2}}{2}\right) _{t}+\func{div}%
\left( \left( \varepsilon +\frac{\left\vert \mathbf{j}\right\vert ^{2}}{%
2\rho }+P\right) \mathbf{u}+c\left( \mathbf{w}\func{div}\ \mathbf{j-j}\,%
\func{div}\ \mathbf{w}\right) \right) =0
\end{equation}%
Since

$\func{div}\left( c\mathbf{ww}^{\ast }-c\left( \rho \func{div}\ \mathbf{w}+%
\displaystyle\frac{c\left\vert \mathbf{w}\right\vert ^{2}}{2}I\right)
\right) $

$=c\mathbf{w}^{\ast }\func{div}\ \mathbf{w}+c\mathbf{w}^{\ast }\displaystyle%
\left( \frac{\partial \mathbf{w}}{\partial \mathbf{x}}\right) ^{\ast }-c%
\mathbf{w}^{\ast }\func{div}\ \mathbf{w-}c\rho \mathbf{\nabla }\left( \func{%
div}\ \mathbf{w}\right) -c\mathbf{w}^{\ast }\displaystyle\left( \frac{%
\partial \mathbf{w}}{\partial \mathbf{x}}\right) $

$=\mathbf{-}c\rho \mathbf{\nabla }\left( \func{div}\ \mathbf{w}\right) $%
\newline
\newline
the momentum equation reads
\[
\mathbf{j}^{\ast}_{\,t}+\func{div}\left( \displaystyle\frac{\mathbf{jj}%
^{\ast }}{\rho }+P\,I\right) \mathbf{-}c\rho \mathbf{\nabla }\left( \func{div%
}\ \mathbf{w}\right) \newline
=\mathbf{0}^{\ast }
\]%
The gradient of the mass conservation law verifies another conservation law%
\[
\mathbf{w}_{t}^{\ast }+\mathbf{\nabla }\left( \func{div}\ \mathbf{j}\right)
=0
\]
If we add an initial condition such that
\[
\mathbf{w}^{\ast }|_{\,t=0}=\mathbf{\nabla }\rho |_{\,t=0}
\]%
we can consider $\mathbf{w}$ as an independent variable. The fact that $%
\mathbf{w}^{\ast }=\mathbf{\nabla }\rho $ will be a consequence of the
governing equations.\ Finally, we obtain equations (6) in the following
equivalent non-divergence form
\begin{equation}
\begin{array}{c}
\rho _{t}+\func{div}\ \mathbf{j}=0 \\
\eta _{t}+\func{div}\displaystyle\frac{\eta }{\rho }\ \mathbf{j}=0 \\
\mathbf{j}_{\,t}^{\ast }+\func{div}\left( \displaystyle\frac{\mathbf{jj}%
^{\ast }}{\rho }+P\,I\right) \mathbf{-}c\rho \mathbf{\nabla }\left( \func{div%
}\ \mathbf{w}\right) \newline
=\mathbf{0}^{\ast } \\
\mathbf{w}_{t}^{\ast }+\mathbf{\nabla }\left( \func{div}\ \mathbf{j}\right)
=0%
\end{array}%
\end{equation}%
The theory of capillary fluids is usually applied for van der Waals-like
fluids. For such fluids the energy $\varepsilon \left( \rho ,\eta \right) $
is not convex for all values of $\rho $ and $\eta $. We suppose that we are
in the vicinity of an equilibrium state $\left( \rho _{e},\eta _{e}\right) $
where the energy function is locally convex.\ Let us introduce conjugate
variables $(q,\ \theta ,\ \mathbf{u},\ \mathbf{r)\ }$by the formula
\begin{equation}
\begin{array}{c}
dE\equiv \displaystyle d\left( \varepsilon +\frac{\left\vert \mathbf{j}%
\right\vert ^{2}}{2\rho }+c\frac{\left\vert \mathbf{w}\right\vert ^{2}}{2}%
\right) \equiv \left( \mu -\frac{\left\vert \mathbf{u}\right\vert ^{2}}{2}%
\right) d\rho +\theta d\eta +\mathbf{u}^{\ast }d\mathbf{j+}c\ \mathbf{w}%
^{\ast }d\mathbf{w} \\
\equiv \displaystyle qd\rho +\theta d\eta +\mathbf{u}^{\ast }d\mathbf{j+r}%
^{\ast }d\mathbf{w}%
\end{array}%
\end{equation}%
The Lagrange transformation of the total energy $E$ is defined by
\[
\Pi =\rho q+\eta \theta +\mathbf{j}^{\ast }\mathbf{u}+\mathbf{w}^{\ast }%
\mathbf{r}-E=P+\frac{\left\vert \mathbf{r}\right\vert ^{2}}{2c}
\]%
where the thermodynamic pressure $P$ is considered as a function of $q,\
\theta $ and $\mathbf{u}$. Hence, in terms of the conjugate variables $(q,\
\theta ,\ \mathbf{u},\ \mathbf{r)}$ defined by equation (9), equations (8)
can be rewritten in the following form
\begin{equation}
\begin{array}{c}
\displaystyle\left( \frac{\partial \Pi }{\partial q}\right) _{t}+\func{div}%
\left( \frac{\partial (\Pi \mathbf{u)}}{\partial q}\right) =0 \\
\displaystyle\left( \frac{\partial \Pi }{\partial \theta }\right) _{t}+\func{%
div}\left( \frac{\partial (\Pi \mathbf{u)}}{\partial \theta }\right) =0 \\
\displaystyle\quad \left( \frac{\partial \Pi }{\partial \mathbf{u}}\right)
_{t}+\func{div}\left( \frac{\partial (\Pi \mathbf{u)}}{\partial \mathbf{u}}-
\frac{\partial \Pi }{\partial q}\dfrac{\partial \mathbf{r}}{\partial \mathbf{%
x}}\right) =0 \\
\displaystyle\quad \left( \frac{\partial \Pi }{\partial \mathbf{r}}\right)
_{t}+\func{div}\left( \frac{\partial (\Pi \mathbf{u)}}{\partial \mathbf{r}}+
\frac{\partial \Pi }{\partial q}\dfrac{\partial \mathbf{u}}{\partial \mathbf{%
x}}\right) =0%
\end{array}%
\end{equation}

\noindent If the capillary coefficient $c$ is zero, $\Pi =P$ and we get the
gas dynamics equation and the symmetric form of Godunov (1961).

Multiplying equations (10) by $q,\ \theta ,~\mathbf{u}$ and $\mathbf{r}$%
\textbf{, }summing up all of them and using the identity
\[
\text{rot}\ \left( \mathbf{a}\times \mathbf{b}\right) =\ \left[ \mathbf{a},%
\mathbf{b}\right] +\mathbf{a}\func{div}\mathbf{b}-\mathbf{b}\func{div}%
\mathbf{a}
\]%
where
\[
\left[ \mathbf{a},\mathbf{b}\right] =\frac{\partial \mathbf{a}}{\partial
\mathbf{x}}\,\mathbf{b}-\frac{\partial \mathbf{b}}{\partial \mathbf{x}}\,%
\mathbf{a}
\]%
denotes the Poisson bracket,
we get the conservation of the energy (7) in the form%
\[
\left( q\frac{\partial \Pi }{\partial q}+\theta \frac{\partial \Pi }{%
\partial \theta }+\frac{\partial \Pi }{\partial \mathbf{u}}\,\mathbf{u+}%
\frac{\partial \Pi }{\partial \mathbf{r}}\,\mathbf{r}-\Pi \right) _{t}
\]%
\[
+\func{div}\left( q\frac{\partial \left( \Pi \mathbf{u}\right) }{\partial q}%
+\theta \frac{\partial \left( \Pi \mathbf{u}\right) }{\partial \theta }+%
\frac{\partial \left( \Pi \mathbf{u}\right) }{\partial \mathbf{u}}\,\mathbf{%
u+}\frac{\partial \left( \Pi \mathbf{u}\right) }{\partial \mathbf{r}}\,%
\mathbf{r-}\Pi \mathbf{u}\right.
\]%
\[
+\left. \frac{\partial \Pi }{\partial q}\,\frac{\partial \mathbf{u}}{%
\partial \mathbf{x}}\mathbf{r-}\frac{\partial \Pi }{\partial q}\,\frac{%
\partial \mathbf{r}}{\partial \mathbf{x}}\mathbf{u}\right) \mathbf{=0}
\]%
The system (10) admits constant solutions $(\rho _{e},\eta _{e},\mathbf{u}%
_{e},\mathbf{w}_{e}=\mathbf{0})$.\ Since the governing equations are
invariant under Galilean transformation, we can assume that $\mathbf{u}_{e}=%
\mathbf{0}$.\ If we look for the solution of the linearized system
proportional to $\displaystyle e^{i\left( \mathbf{k}^{\ast }\mathbf{x}%
-\lambda t\right) }$, we get equation (3), in which we have put
\[
\begin{array}{c}
\displaystyle A=\frac{\partial }{\partial \mathbf{v}}\left( \left( \frac{%
\partial \Pi }{\partial \mathbf{v}}\right) ^{\ast }\right) ,\quad \quad
B=\sum_{i=1}^{n}B^{i}k_{i},\quad \quad B^{i}=\frac{\partial }{\partial
\mathbf{v}}\left( \left( \frac{\partial \Pi u^{i}}{\partial \mathbf{v}}%
\right) ^{\ast }\right) , \\
C=-C^{\ast }=\left(
\begin{array}{ccc}
0 & \mathbf{0}^{\ast } & \mathbf{0}^{\ast } \\
\mathbf{0} & \mathit{O} & -\rho _{e}\mathbf{kk}^{\ast } \\
\mathbf{0} & \rho _{e}\mathbf{kk}^{\ast } & O%
\end{array}%
\right) \quad \text{with}\quad \mathit{O}=\left(
\begin{array}{ccc}
0 & 0 & 0 \\
0 & 0 & 0 \\
0 & 0 & 0%
\end{array}%
\right)
\end{array}%
\]%
and $\mathbf{v}^{\ast }=\left( q,\ \theta ,~\mathbf{u}^{\ast },\ \mathbf{r}%
^{\ast }\right) $.\ Hence, eigenvalues $\lambda $ are real if $A$ is
positive definite.

\section{One-dimensional barotropic case}

In mass Lagrangian coordinates $(t,z)$ the governing equations are (see
Gavrilyuk and Serre, 1995)%
\[
v_{t}-u_{z}=0,\ \ \ u_{t}+p_{z}=0,\ \ \ p=-\frac{\delta e}{\delta v}=-\left(
\frac{\partial e}{\partial v}-\frac{\partial }{\partial z}\left( \frac{%
\partial e}{\partial v_{z}}\right) \right) ,\ \ \ e=e(v,v_{z})
\]%
where $v=\displaystyle\frac{1}{\rho} $ denotes the specific volume.\ This
case is general: we do not suppose a particular form (4) of the energy $e$.\
Consider an augmented system
\begin{equation}
\begin{array}{c}
v_{t}-u_{z}=0 \\
w_{t}-u_{zz}=0 \\
{u_{t}-\left( \displaystyle\frac{\partial e}{\partial v}-\frac{\partial }{%
\partial z}\left( \frac{\partial e}{\partial w}\right) _{z}\right) _{z}=0}%
\end{array}%
\end{equation}%
Let us define $\pi $ and the conjugate variables $(\sigma ,r)$ as%
\[
\pi =\frac{\partial e}{\partial v}\, v+\frac{\partial e}{\partial w}%
\,w\equiv \sigma v+rw-e
\]

\bigskip

\noindent In terms of $\pi $\ and $(\sigma ,r,u)$ the system (11) reads
\[
\begin{array}{c}
\left( \displaystyle\frac{\partial \pi }{\partial \sigma }\right)
_{t}-u_{z}=0 \\
\left( \displaystyle\frac{\partial \pi }{\partial r}\right) _{t}-u_{zz}=0 \\
u_{t}-\left( \displaystyle\sigma -r_{z}\right) _{z}=0%
\end{array}
\]

\bigskip

\noindent In matrix form we get
\begin{equation}
A\ \left(
\begin{array}{c}
\sigma \\
r \\
u%
\end{array}%
\right) _{t}+B_{1}\ \left(
\begin{array}{c}
\sigma \\
r \\
u%
\end{array}%
\right) _{z}+C_{1}\left(
\begin{array}{c}
\sigma \\
r \\
u%
\end{array}%
\right) _{zz}=0
\end{equation}%
where
\begin{equation}
A=\left(
\begin{array}{ccc}
\displaystyle\frac{\partial ^{2}\pi }{\partial \sigma ^{2}} & \displaystyle%
\frac{\partial ^{2}\pi }{\partial \sigma \partial r} & 0 \\
&  &  \\
\displaystyle\frac{\partial ^{2}\pi }{\partial \sigma \partial r} & %
\displaystyle\frac{\partial ^{2}\pi }{\partial r^{2}} & 0 \\
&  &  \\
0 & 0 & 1%
\end{array}%
\right) ,\ B_{1}=\left(
\begin{array}{ccc}
0 & 0 & -1 \\
0 & 0 & 0 \\
-1 & 0 & 0%
\end{array}%
\right) ,\ C_{1\ }=\left(
\begin{array}{ccc}
0 & 0 & 0 \\
0 & 0 & -1 \\
0 & 1 & 0%
\end{array}%
\right)
\end{equation}

\bigskip

\bigskip

\noindent Equations (12) and (13) imply a dispersion relation of type (3),
if we put $B=kB_{1}$, $C=k^{2}C_{1}$.\ We note also that the system admits
the energy conservation law%
\[
\left( \frac{u^{2}}{2}-\sigma \frac{\partial \pi }{\partial \sigma }+r\frac{%
\partial \pi }{\partial r}-\pi \right) _{t}+\Big(-\sigma u+\left[ r,u\right]%
\Big)_{z} =0
\]%
where $\left[ r,u\right] =ur_{z}-ru_{z}.$ \bigskip

\bigskip

\noindent \textbf{Remark.} Analogous symmetric forms may be obtained for
bubbly liquids, where the internal energy is a function not only of the
density but also of the total derivative of the density with respect to time.

\end{document}